\documentclass[article]{aa} 
\usepackage{txfonts}
\usepackage{natbib}
\usepackage{graphicx}
\usepackage{color}
\usepackage{longtable,lscape}
\usepackage{float}
\usepackage{subfigure}%

\newcommand{\kmpers}{$\mathrm{km \, s^{-1}}$} % units of velocity

\newcommand{\cmthree}{cm$^{-3}$}

\newcommand{\vlsr}{$\upsilon_{\rm LSR}$}              %velocities

\newcommand{\about}{$\sim$}                       %approx
\newcommand{\expo}[1]{$10^{#1}$}
\newcommand{\texpo}[1]{$\,\times\,10^{#1}$}

\newcommand{\nhtva}{$n$($\mathrm{H_2}$)}           % densitites

\newcommand{\htvao}{H$_2$O}  
\newcommand{\htva}{H$_2$}

\newcommand{\htvaartonotva}{$\rm{H_2^{18}O\,(4_{14}-3_{21})}$}
\newcommand{\hdoett}{$\rm{HDO\,(1_{01}-0_{00})}$}
\newcommand{\asec}{$^{\prime \prime}$}
\newcommand{\adeg}{$^{\circ}$}

\newcommand{\atwozero}{$\alpha_{2000}$}
\newcommand{\dtwozero}{$\delta_{2000}$}

\newcommand{\radot}[4]{\mbox{#1$^{\rm h}$#2$^{\rm m}$#3$\stackrel{^{\rm
s}}{_{\bf\cdot}}$#4}}
\newcommand{\decdot}[3]{\mbox{#1$^{\circ}$#2$^{\prime}$#3$^{\prime \prime}$}}
\newcommand{\lsun}{$L_{\odot}$}                          %solar and terr. units
\newcommand{\lbol}{$L_{\rm{bol}}$}

\newcommand{\radex}{\texttt{RADEX} }            % programs

\newcommand{\irasfemtontre}{\mbox{IRAS\,15398--3359}}

\newcommand{\jorgensen}{J\o rgensen}

\authorrunning{P. Bjerkeli, et al.}
\titlerunning{Water around \irasfemtontre\ as observed with ALMA}

\begin{document}

   \title{\textbf{\Large{Water around \irasfemtontre\ observed with ALMA
   %\thanks{Thanks....}
   }}}

   %%\subtitle{}

   \author{P. Bjerkeli
          \inst{1,2},
          J.~K. J\o rgensen\inst{1}, 
          E.~A. Bergin\inst{3}, S. Frimann\inst{1}, D. Harsono\inst{4}, S.~K. Jacobsen\inst{1}, \\ J.~E. Lindberg\inst{5}, M. Persson\inst{4}, N. Sakai\inst{6}, E.~F. van~Dishoeck\inst{4,7}, R. Visser\inst{3}, S. Yamamoto\inst{8}    
          }
   \institute{
             %Niels Bohr Institute, University of Copenhagen, Juliane Maries Vej 30, DK-2100 Copenhagen \O, Denmark \\
             Centre for Star and Planet Formation, Niels Bohr Institute \& Natural History Museum of Denmark, University of Copenhagen, {\O}ster Voldgade 5--7, 1350 Copenhagen K., Denmark \\
             \email{per.bjerkeli@nbi.dk}
             %\and
             %Centre for Star and Planet Formation and Natural History Museum of Denmark, University of Copenhagen, \O ster Voldgade 5--7, DK-1350 Copenhagen K, Denmark 
             \and
             Department of Earth and Space Sciences, Chalmers University of Technology, Onsala Space Observatory, 439 92 Onsala, Sweden 
             \and
             Department of Astronomy, University of Michigan, 1085 S. University Ave., Ann Arbor, MI 48109-1107, USA
             \and
             Leiden Observatory, Leiden University, Niels Bohrweg 2, 2333 CA, Leiden, the Netherlands
             \and
              NASA Goddard Space Flight Center, Astrochemistry Laboratory, Mail Code 691, 8800 Greenbelt Road, Greenbelt, MD 20771, USA
             \and
             The Institute of Physical and Chemical Research (RIKEN), 2-1, Hirosawa, Wako-shi, Saitama 351-0198, Japan
             \and
             Max-Planck-Institut f\"ur Extraterrestische Physik, Giessenbachstrasse 2, 85478, Garching, Germany
             \and
             Department of Physics, Graduate School of Science, The University of Tokyo, 7-3-1 Hongo, Bunkyo, Tokyo 113-0033, Japan       
          %   \vspace{2.5cm}\\
                          }

   \date{Received 26 April 2016 / Accepted 1 August 2016}

\abstract
{Understanding how protostars accrete their mass is one of the
  fundamental problems of star formation. High dust column
  densities and complex kinematical structures make direct observations challenging. Moreover, \textcolor[rgb]{1,0.501961,0}{ \textcolor[rgb]{0,0,0}{direct observations}} {}only provide a  snapshot. Chemical tracers provide an interesting alternative to
  characterise the infall histories of protostars.}
   {We aim to map the distribution and kinematics of
     gaseous water towards the low-mass embedded protostar \irasfemtontre. Previous observations of
     H$^{13}$CO$^+$ showed a depression in the abundance towards
     \irasfemtontre.\  This is a sign of destruction of HCO$^+$ by an
     enhanced presence of gaseous water in an extended region, possibly related
     to a recent burst in the accretion. Direct observations of water
     vapour can determine the exact extent of the emission and 
     confirm the hypothesis that HCO$^+$ is indeed a good tracer of
     the water snow-line.}
   {\irasfemtontre\ was observed using the Atacama Large
     Millimeter/submillimeter Array (ALMA) at 0.5$''$ resolution in
     two setups at 390 and 460~GHz. Maps of \hdoett\ and
     \htvaartonotva\ were taken simultaneously with observations of the
     CS\,(8--7) and N$_2$H$^+$\,(5--4) lines and continuum at 0.65
     and 0.75~mm. The maps were interpreted using dust radiative
     transfer calculations of the protostellar infalling envelope with
     an outflow cavity.}
   {HDO is clearly detected and extended over the scales of the
     H$^{13}$CO$^+$ depression, although it is displaced
by \about500~AU in the
     direction of the outflow. H$_2^{18}$O is tentatively detected towards the red-shifted outflow lobe, but otherwise it is absent from the mapped region, which suggests that temperatures are low. Although we cannot entirely exclude a shock origin, this indicates that another process is
responsible for the water emission.}
   {Based on the temperature structure obtained from dust radiative transfer models, we conclude that the
     water was most likely released from the grains in an extended
     hour-glass configuration during a recent accretion burst. HDO is
     only detected in the region closest to the protostar, at
     distances of up to 500 AU. These signatures can only be explained if
     the luminosity has recently been increased by orders of
     magnitudes. Additionally, the densities in the
     outflow cones must be sufficiently low. }


\maketitle
\section{Introduction}
\label{section:introduction}
When stars form, the bulk of the luminosity is due to the
accretion of material onto the protostellar object. It remains a
major challenge, however, to directly probe the material as it falls in from the
natal cloud. While outflows have been shown to be prominent
\citep{Snell:1980lr}, the accretion process is heavily obscured by the
dense environments close to protostars. Understanding this process is fundamentally important, however, because it has profound implications on
the properties of stars, the initial mass function, and the physical
and chemical composition
of circumstellar material. In recent years, several studies have focused 
on the question whether accretion
takes place at a constant rate or if violent bursts occur in the
accretion rate. Accretion bursts might be the result of
fragmentation of protostellar disks \citep[see
e.g.][]{Bell:1994rc,Vorobyov:2006yg,Vorobyov:2013qf} or be related to the
nature of the large-scale infall \citep{Padoan:2014mz}. Variations
like this are also tied to the so-called luminosity problem
\citep{Kenyon:1990bh}, where the observed current luminosities of protostars
generally are more than an order of magnitude lower than inferred from standard accretion models \citep[see e.g.][]{Evans:2009rt}. It is therefore possible that protostars spend most of their lifetimes in a quiescent phase and accordingly seem to be less luminous than expected. Bursts in accretion are known to be common and long-lasting in the more evolved FU Orionis stars \citep{Herbig:1977qy}, but it still remains a challenge to determine whether this phenomenon is equally common among younger objects.

Large variations in the accretion rates of protostars may also have
important consequences for their chemical structures. To first order,
the chemistry is strongly affected by the location where molecules are frozen 
out on and sublimate from dust grains at the so-called
snow lines. For a protostar of constant luminosity, the location of
these boundaries for the chemistry is well-defined based on how large
a region the central protostar can heat up to and above the
sublimation temperature of the molecule. However, when a protostar
undergoes a burst and significantly increases its luminosity, the dust grains will heat up almost instantly on
larger scales \citep{Johnstone:2013oq}, allowing molecules to
sublimate from the grain ice-mantles on timescales shorter than a few years \citep[e.g.][]{Rodgers:2003fj}. The opposite process,
freeze-out, generally takes a much longer time at the
densities characteristic of protostellar envelopes, however: typically \expo{4} -- \expo{5} years for CO and \htvao\ at the density \nhtva~=~\expo{5}~\cmthree\ and 100 -- 1000 years at \nhtva~=~\expo{7}~\cmthree\ \citep[e.g.][]{Rodgers:2003fj}. This suggests that
observations of even relatively common molecules may provide an
indirect way of estimating whether a given protostar has recently
undergone a burst in accretion
\citep[e.g.][]{Visser:2012yu,Vorobyov:2013qf,Visser:2015jk}.

Recent observations \citep{Jorgensen:2015yu} of a sample of 16 embedded
protostars with the SMA show that the extent of the C$^{18}$O emission varies
significantly and is not directly correlated with the current
luminosities of the protostars. About half the sources show
extended CO emission that would imply that the protostars have
undergone bursts within the past 10$^4$~years. Freeze-out onto the grains occurs for CO whenever the temperature drops below
\about30~K \citep{Noble:2012yu}, which typically occurs on relatively large scales in the
protostellar envelopes where the density is low and the
timescale for freeze-out accordingly is relatively long. To find the sources with the most recent
bursts, snow-line tracers at higher densities closer to the
protostars are required. Molecules with snow lines at high densities can also trace 
bursts of higher intensities.

\irasfemtontre\ \citep{Shirley:2000qf} is a Class 0
\citep{Andre:1990fk,Andre:1993fk} protostar located in the Lupus I
cloud at a distance of 155 pc \citep{Lombardi:2008lr}. The source has
a prominent bipolar molecular outflow
\citep{Tachihara:1996vn,van-Kempen:2009rt,Dunham:2014fk,Yildiz:2015kq,Jorgensen:2013lr,Oya:2014kx}
that extends less than 3000 AU from the protostar itself,
possibly indicating that the source is very young
\citep{Bjerkeli:2016lr}. The bolometric luminosity is meassured
at 1.8~\lsun\ \citep{Jorgensen:2013lr}, and the source shows a
very interesting chemistry. It is one of two known so-called warm
carbon-chain chemistry sources \citep{Sakai:2009qq,Sakai:2011fr}.
\irasfemtontre\ was observed with the Atacama Large
Millimeter/submillimeter Array (ALMA) in Cycle 0, and these
observations revealed a hole in the H$^{13}$CO$^+$\,(3--2) emission
\citep{Jorgensen:2013lr}, while other species (i.e. C$^{17}$O and
CH$_3$OH) peak at the central position. As discussed by these
authors, the best explanation is an increased abundance of
gaseous water due to a recent accretion burst occurring 100 -- 1000 years ago. In this scenario, the abundance of gaseous water is expected to increase because of the increased stellar luminosity, and the time estimate was based on the expected freeze-out time for
water. The abundance of HCO$^+$ is expected to decrease whenever water
is abundant and temperatures and densities are high through the reactions\begin{eqnarray*}
& \rm{CO + H_3^+} & \rightarrow \rm{HCO^+ + H_2}, \nonumber \\
& \rm{HCO^+ + H_2O} & \rightarrow \rm{H_3O^+ + CO}. \nonumber
\end{eqnarray*}
The most surprising aspect of the H$^{13}$CO$^+$ map
presented in \citet{Jorgensen:2013lr} is, however, that the extent of
the hole (150 -- 200~AU) is much larger than the region where
temperatures are expected to be higher than the sublimation temperature 
for water, that is, at 100~K. Given the current luminosity of the protostar, the temperature 
decreases to below 100~K as close as \about20~AU (see Sect.~\ref{sec:analysisanddiscussion}) 
from the protostar.

To explain the
extent of the H$^{13}$CO$^+$ absence (and the presence of water), the
protostar must have reached a luminosity one to two orders of
magnitude higher than its current value. This conclusion is supported by 
the CO map obtained with the SMA, which reveals dynamical events in
the outflow on timescales of a few hundred years
\citep{Bjerkeli:2016lr}. An increased accretion rate would probably
result in an increased outflow activity during the period of the burst
\citep{Raga:1990ij}. However, to confirm this hypothesis, observations
of the water molecule itself are required. Since the atmosphere is
opaque to the lowest-energy transitions from H$_2^{16}$O and because
the spatial resolution available with space-based facilities
is poor, we
here used isotopologues that can be readily observed from
the ground. In this program, which was carried out using ALMA during
Cycle~2, two isotopologue transitions were targeted to map the
distribution of thermal water: \htvaartonotva\ and \hdoett. These
two transitions are complementary in the sense that they probe gas
at very different excitation (\mbox{$E_{\rm{up}}=390~$K} and \mbox{$E_{\rm{up}}=22~$K} ). Whether both HDO and H$_2^{18}$O or only
HDO is detected therefore depends on the thermal
structure of the region.

We discuss the observations of H$_2^{18}$O and HDO in
the vicinity of \irasfemtontre.  We also discuss how the location of
the snow line might be affected by the molecular outflow known to be
present in the region. The observations and results are presented
in Sects.~2 and 3. The emission from water is discussed in Sect.~4, and
the results are summarised in Sect.~5.

\begin{figure*}[]
   \flushleft
   %\vspace{-1cm}
   %\hspace{-1.10cm}
    \rotatebox{0}{\includegraphics[width=1.10\textwidth]{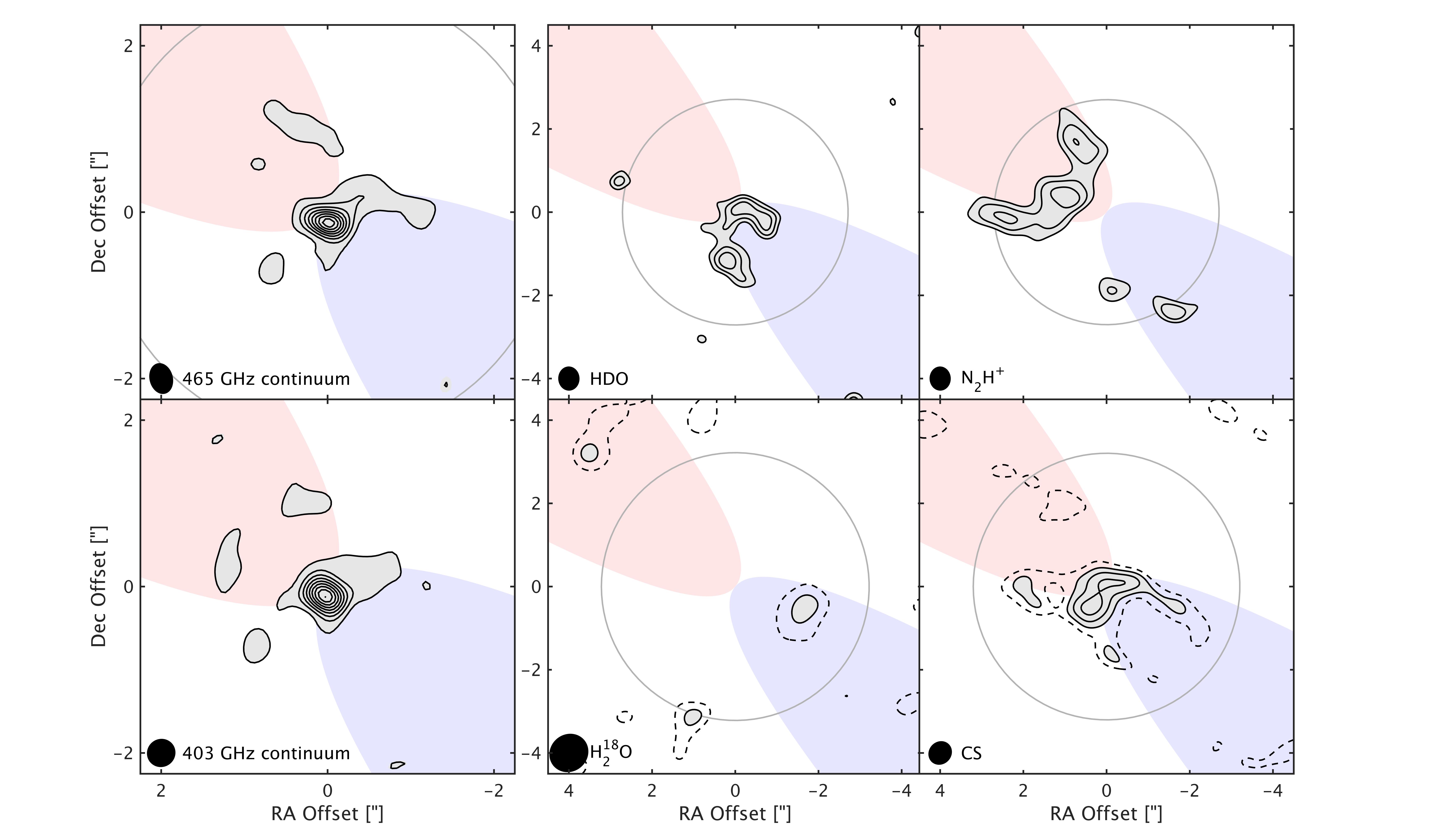}}
     % \vspace{-0.7cm}
      \caption{Continuum and integrated line emission maps towards \irasfemtontre, centred on \atwozero~=~\radot{15}{43}{02}{24}, \dtwozero~=~\decdot{--34}{09}{06.7}. The lines are integrated over a velocity range of $\Delta \upsilon$~=~0.75~\kmpers\ (except for H$_2^{18}$O, where $\Delta \upsilon$~=~2.00~\kmpers\ has been used). Contours are from 3$\sigma$ in steps of 1$\sigma$. For H$_2^{18}$O and CS, the 2$\sigma$ level is also indicated with dashed contours. The red and blue shaded regions mark the directions of the outflow cones (same geometry as in the models discussed in Sect.~\ref{section:icesublimationduringarecentaccretionburst}). The 90-percent level of the beam profile is indicated with a circle in each panel. The beam is shown in the lower left corner of each map. The continuum maps are zoomed-in by a factor of two.  }
         \label{fig:speciesmaps}
   \end{figure*}
\section{Observations}
\label{section:observations}
\irasfemtontre\ was observed using ALMA on 2015 May 20 and 21
as part of the Cycle~2 program 2013.1.00244.S (PI: Jes
\jorgensen). The main science goal was to observe the \hdoett\ line at
464.925~GHz and the \htvaartonotva\ line at 390.608~GHz. Thirty-six antennas
in the 12 m array were used during the observations, providing
baselines in the range 20 m ($\approx$25--30 k$\lambda$) to 520 m
($\approx$700--800 k$\lambda$). The project was carried out under
excellent observing conditions with a precipitable water vapour ranging
between 0.26~mm and 0.41~mm. The phase centre was at
\atwozero~=~\radot{15}{43}{02}{24},
\dtwozero~=~\decdot{--34}{09}{06.7} and the total observing time was
3.6 hours. Two different settings were used to cover both the 
HDO and the H$_2^{18}$O line. The HDO setup contains four line
spectral windows centred on 462.607~GHz, 463.719~GHz, 464.928~GHz, and
465.818~GHz, as well as two continuum spectral windows centred on
463.104~GHz and 465.000~GHz. The bandwidth for the line spectral
windows is 234~MHz, and for the continuum spectral windows it
is 2000~MHz. The corresponding channel spacings are 0.122~MHz and
15.625~MHz, respectively. The H$_2^{18}$O observation contains three
line spectral windows centred on 388.616~GHz, 390.612~GHz, and
391.851~GHz, and two continuum spectral windows centred on 401.104~GHz
and 402.904~GHz. The bandwidths and channel spacings are the same as
for the HDO setup. In the settings described above, N$_2$H$^+$\,(5--4), \mbox{CS\,(8--7)}, and \mbox{C$^{33}$S\,(8--7)} were covered as well. The observations are summarised in Table~\ref{table:correlator}.
\begin{table*}[t!]
\flushleft
\caption{ALMA observations. }
%\resizebox{\hsize}{!}{
\begin{tabular}{llllllll}\hline\hline
  \noalign{\smallskip}
Spectral window & Number of channels / resolution & Center frequency & Bandwidth & Line & $E_{\rm{up}}$\,(K) \\ 
  \noalign{\smallskip}
\hline
  \noalign{\smallskip}
\multicolumn{3}{l}{\textit{HDO setting: }} \\
  \noalign{\smallskip}
  0 & 960 / 244.141 kHz & 465.818 GHz & 234.375 MHz & N$_2$H$^+$\,(5--4) & 67.1\\
  1 & 960 / 244.141 kHz & 464.928 GHz & 234.375 MHz & \hdoett\ & 22.3\\
  2 & 128 / 15625.000 kHz & 465.000 GHz & 2000.000 MHz & Continuum & - \\
 \noalign{\smallskip}
\multicolumn{3}{l}{\textit{H$_2^{18}$O setting:} } \\
 \noalign{\smallskip}   
 0 & 960 / 244.141 kHz & 390.612 GHz & 234.375 MHz & \htvaartonotva\ & 322.0 \\
 1 & 960 / 244.141 kHz & 391.851 GHz & 234.375 MHz & CS\,(8--7) & 84.6 \\
 4 & 128 / 15625.000 kHz & 402.904 GHz & 2000.000 MHz & Continuum & - \\
  \noalign{\smallskip}  
\hline
\end{tabular}
\label{table:correlator}
%}
\end{table*}

The data reduction was carried out in CASA \citep{McMullin:2007nr} and
followed the standard procedure. The subsequent analysis of the maps
was made in MATLAB. The phase calibration was carried out through
observations of the BL Lac object J1517-2422, bandpass calibration on
the quasar J1256-0547, and flux calibration on Titan. One antenna
(DA42) was flagged because of an uncertainty in the position. This has no
significant effect on the reduced image cubes and does not alter any
of the conclusions presented in this paper, nor does it have a
significant effect on the presented maps. The output continuum image cubes
have an angular resolution \about0.3\asec\ , while the line image cubes have a resolution of \about0.5\asec. The difference arises because a uv taper was used when CLEANing the lines.

\section{Results}
The key aspect of this program was to image the emission of the two
water isopologues, HDO and H$_2^{18}$O. The emission maps for these
two, along with the other detected species, are presented in Fig.~\ref{fig:speciesmaps}.  
\label{section:results}   
\subsection{HDO and H$_2^{18}$O}
\begin{figure}[h!]
   \flushleft
   %\vspace{-1cm}
   %\hspace{-1.10cm}
    \rotatebox{0}{\includegraphics[width=0.49\textwidth]{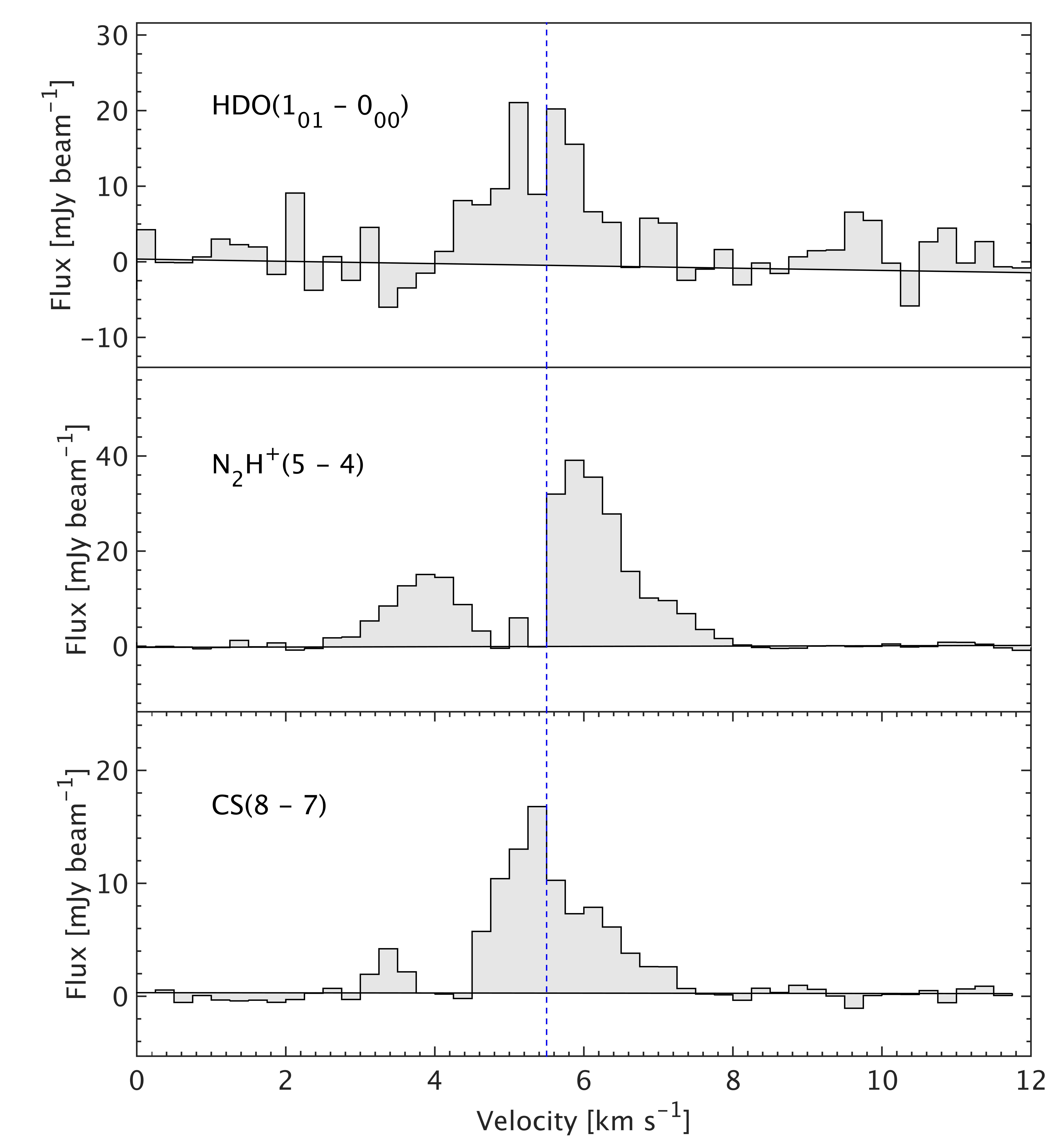}}
     % \vspace{-0.7cm}
      \caption{All spectra within 2\asec\ (1\asec\ for \hdoett) from the centre of the map, averaged together. The vertical dashed line represents the source velocity at \vlsr~=~5.5~\kmpers.}
         \label{fig:spectra}
   \end{figure}

The \hdoett\ line is detected at the 5$\sigma$ level
(integrated flux) in the vicinity of the protostar and in the
blue-shifted part of the outflow (see Fig.~\ref{fig:speciesmaps}). The spectral
location of the line peak intensity is somewhat uncertain because
the signal-to-noise ratio is low, but we estimate it to be in the
range \vlsr~=~5.0 -- 5.5~\kmpers. The line width is narrow with
a FWHM $\simeq$1~\kmpers. In Fig.~\ref{fig:spectra} the
spectra within 2\asec\ of the continuum peak are shown. The HDO spectrum represents
the average of the emission within 1\asec\ from the central position. 

\begin{figure}[h!]
\flushleft
\vspace{0.3cm}
%   \begin{tabular}{lll}
   \vspace{-0.5cm}
      \rotatebox{0}{\includegraphics[width=0.48\textwidth]{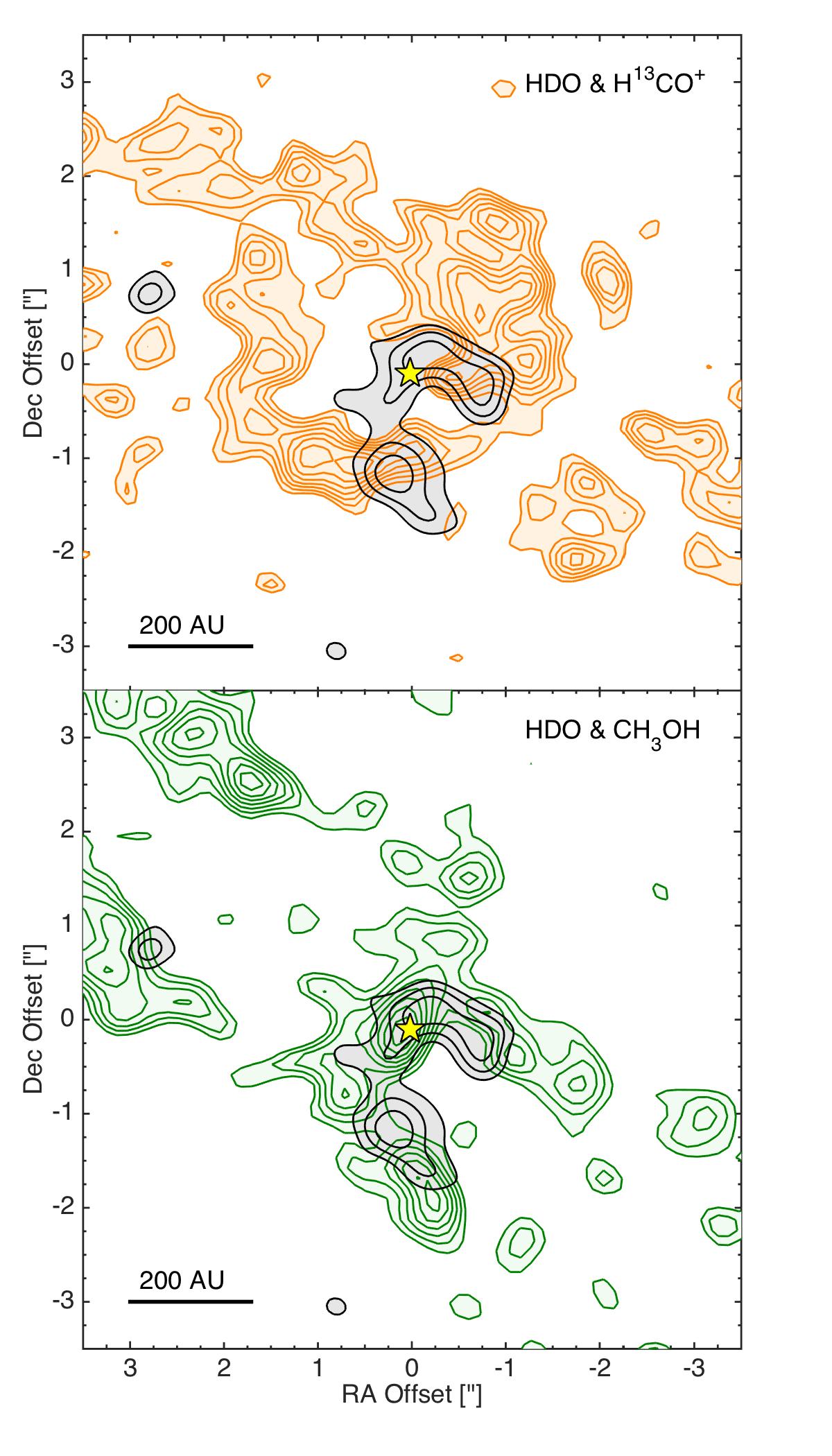}} 
      \caption{HDO emission (black) compared to the H$^{13}$CO$^+$ emission (brown, upper panel) and the CH$_3$OH emission (green, lower panel) presented in \citet{Jorgensen:2013lr}. Contours are from 3$\sigma$ in steps on 1$\sigma$. The continuum peak is indicated with a star.}
         \label{fig:hdoh13co}
   \end{figure}

   HDO is detected in the 
   region where H$^{13}$CO$^+$ is not detected \citep[see Fig.~\ref{fig:hdoh13co}, and][their
   Fig.~1]{Jorgensen:2013lr}, but an extension of the HDO
   emission towards the south-west of the H$^{13}$CO$^+$ ring is visible as
   well. 
     The extent of this emission is \about300~AU and follow the cavity walls of the blue-shifted outflow outlined by the CH$_3$OH emission.  Furthermore, a 
   red-shifted component is observed in the red-shifted outflow
   lobe at the 4$\sigma$ level. This feature is spatially coincident
   with the emission knot observed in high-velocity CO\,(2--1)
   \citep[][their Fig.~3]{Bjerkeli:2016lr} and the separation to the central source 
  is \about500~AU. In Fig.~\ref{fig:hdovel} we plot the 3.5$\sigma$ level contours only for emission integrated over different velocity intervals. Emission at higher velocity with respect to the systemic velocity are indicated in this figure with darker red and blue than the lower velocity emission.  Comparing different velocity bins in this way shows a possible gradient in the mapped region. In general, 
   emission at higher velocities is detected at greater distances
   from the source than at lower velocities.
  
   The \htvaartonotva\ line in contrast to HDO is not detected towards the
   protostar. There are two tentative features in the outflow lobes, but
   only at the 3$\sigma$ level. The north-eastern of these emission
   knots is located downstream of the feature detected in \hdoett.

\subsection{Continuum and other lines}
The continuum is detected with a peak flux of 36$\pm$1~mJy~beam$^{-1}$ at 465~GHz and 23$\pm$1 mJy~beam$^{-1}$ at 403~GHz, towards
\atwozero~=~\radot{15}{43}{02}{24},
\dtwozero~=~\decdot{--34}{09}{06.8} (from a Gaussian fit). This can be compared to the peak flux of
19~mJy~beam$^{-1}$, reported by \citet{Jorgensen:2013lr} at 345~GHz, where the beam size was similar. 
The largest recoverable scales in the observations presented here range from 4 -- 5\asec. We therefore
cannot exclude that emission from the ambient envelope is resolved
out,
and some low-level extended emission is observed. 
The lowest contours of the emission presented in 
Fig.~\ref{fig:speciesmaps} should therefore be considered as unreliable because we lack short 
spacing data. 

\begin{figure*}[t!]
\flushleft
\vspace{0.3cm}
%   \begin{tabular}{lll}
   \vspace{-0.3cm}
       \rotatebox{0}{\includegraphics[width=1.03\textwidth]{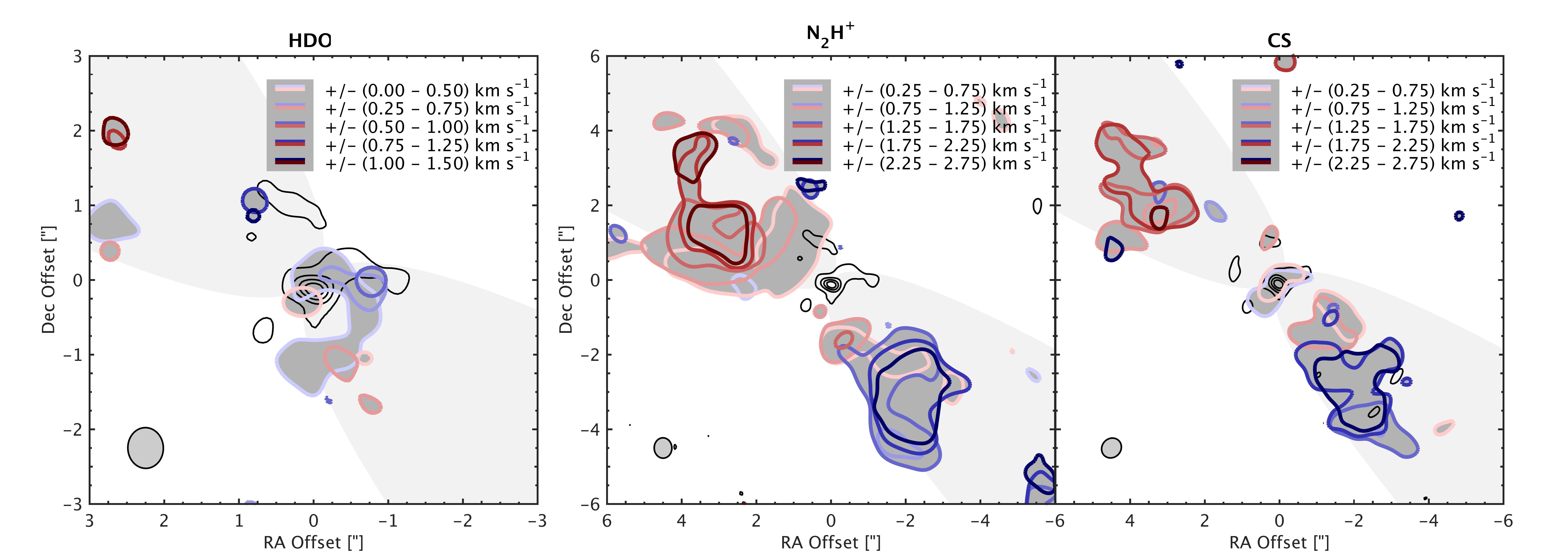}} 
      \caption{3.5$\sigma$ level in different velocity intervals offset from the systemic velocity (see inset). For HDO, the lowest velocities are towards the central region and slightly displaced into the blue-shifted outflow. Higher velocities are detected further out in both outflow lobes. The 465~GHz (for HDO) and 403~GHz (for CS and N$_2$H$^+$) continuum is plotted with black contours. The extent of the outflow is indicated in gray.}
         \label{fig:hdovel}
   \end{figure*}

%\subsection{Other lines}
The N$_2$H$^+$\,(5--4) line at 465.825~GHz is detected towards the
cavity walls in the red-shifted outflow lobe and in a series of knots in
the cavity wall and inner regions of the blue-shifted outflow lobe. No
emission is detected towards the protostar itself. A clear
velocity gradient is visible in this dataset, and higher velocities are
detected farther away from the central source than close to
the outflow launching region (Fig.~\ref{fig:hdovel}). 

CS\,(8--7) is detected towards \irasfemtontre,\ and the
emission shows a flattened, slightly elongated structure. CS, at slightly shifted
velocities (with respect to \vlsr~=~5.5~\kmpers), presents two emission
regions. As for HDO, the red-shifted region is spatially
coincident with the inner emission peak for high-velocity CO\,(2--1)
\citep{Bjerkeli:2016lr}. The same situation applies to the
blue-shifted emission, which is spatially coincident with the high-velocity
CO\,(2--1) blue-shifted knot. This map is presented in
Fig.~\ref{fig:hdovel}, and the emission
reveals a velocity structure towards the protostar that weakly indicates
rotation. The size of this structure is about 200~AU.

The spectra of N$_2$H$^+$ and CS averaged  within 2\asec\ from the
central position are presented in Fig.~\ref{fig:spectra}.
A detailed study of the distribution of the N$_2$H$^+$\,(5--4) and CS\,(8--7)
emission is deferred to a future paper. 

The C$^{33}$S\,(8--7) transition at 391.851~GHz was also covered in the spectral setup, but was not detected at an 
 rms level of \about15~mJy~beam$^{-1}$.

%\section{Analysis}
%\label{section:analysis}
\section{Analysis and discussion}
\label{sec:analysisanddiscussion}
As noted in the previous section, HDO is clearly detected in the inner region and is also displaced into the outflow lobes compared to the centre of the
H$^{13}$CO$^+$ ring, while H$_2^{18}$O is not detected. 
A comparison between H$^{13}$CO$^+$
and HDO is shown in Fig.~\ref{fig:hdoh13co}. Parts of the HDO emission originate in the 
region where H$^{13}$CO$^+$ is not detected, although the extent does not cover the full size of the H$^{13}$CO$^+$
hole. On the other hand, HDO is also abundant in the outflow region where
H$^{13}$CO$^+$ is detected. The low-level HDO emission is not present in the red-shifted outflow close to the central protostar because the short-wavelength dust emission is optically thick and naturally blocks the red-shifted emission at the far side of the protostar. HDO is detected in the blue-shifted outflow lobe while H$^{13}$CO$^+$ is not; this may  be a geometrical effect. HCO$^+$ is known to trace the outflow cavity walls \citep{Bjerkeli:2016lr}, and we cannot exclude from the maps alone that HDO might be tracing a region inside of the HCO$^+$ emission. The fact that HDO is present 
towards the continuum peak supports the scenario outlined by  \citet{Jorgensen:2013lr}.
However, the apparent association with the outflow 
lobes observed here needs an explanation as well. There may be
several causes of this spatial shift and the origin of the HDO emission, 
as we discuss in the sections below.  

\subsection{Water chemistry in shocked outflowing gas?}
Since the HDO emission is spatially coincident with the outflowing gas, it is reasonable to assume that high-temperature chemistry plays an important role in this case.  At high temperatures and densities, neutral-neutral reactions in the gas phase are efficient in forming water, but frozen water can also easily be desorbed from grains \citep[see e.g.][for a recent review on the topic]{van-Dishoeck:2014xe}. These are conditions that are known to be present in molecular outflows and shocks \citep[see e.g.][]{Tafalla:2013qy}. 

%However, there are in this case facts pointing against the scenario where outflow chemistry is solely the cause of the water emission.  
Using \radex\ and assuming relative abundances with respect to \htva\ of $X$(HDO)~=~2\texpo{-7} \citep{Persson:2014oz} and $X$(H$_2^{18}$O)~=~1\texpo{-7} \citep{Wilson:1994qy}, 
we estimate that the gas thermal pressure, $nT$, needs to be as high as \expo{11}~\cmthree~K  to excite the \htvaartonotva\ line at the 3-sigma level, given the strength of the \hdoett\ line. The reason for this is the different upper state energy of the lines (i.e. HDO
at 22 K and H$_2^{18}$O at 390 K), which essentially constrains the temperature of the gas. High densities and temperatures can be found in shocks and in hot corinos, but they are less likely inside the outflow cones. The non-detection of H$_2^{18}$O therefore tells us that the pressure cannot be too high, not even close to the protostar itself. However, it does not exclude a shock origin. Previous studies suggested that \htva\ densities are indeed high in shocked regions, although generally not above \expo{8}~\cmthree\ \citep{Mottram:2014yq}. 

An even stronger argument against the shock scenario is instead the narrow width of the HDO line ($\Delta \upsilon\,\simeq\,$1~\kmpers). In shocks, much higher velocities are expected for water \citep[see e.g.][]{Kristensen:2012kx}. Even if the low-velocity HDO is due to sputtering from the ice \citep{Suutarinen:2014zr} and not to gas-phase reactions in the post-shock gas where the HD/\htva\ abundance can be significantly decreased \citep{Neufeld:2006uq}, we would not expect the lines to be as narrow as \about1~\kmpers. Moreover, a shock origin does not explain why HDO is only detected in the region closest to the protostar. If shock chemistry were the sole explanation to the emission, then there is no reason why we should not also observe HDO at larger distances.

Downstream of the red-shifted knot (Fig.~\ref{fig:speciesmaps}), a tentative H$_2^{18}$O feature is visible. Although we can at present not confirm that it is real, it prohibits us from drawing firm conclusions about the shock
chemistry in this particular region. From these data we cannot
exclude a very high pressure.
\subsection{Ice sublimation during a recent accretion burst?}
\label{section:icesublimationduringarecentaccretionburst}
Another possible explanation of the distribution and kinematics of the HDO emission is that water was released from the grains during a recent accretion burst \citep[i.e. the scenario outlined in][]{Jorgensen:2013lr}.
If the water release from the grains were due to an increase in luminosity, then how close to the protostar did this occur? As mentioned before, the extent of the HDO emission is several hundred AUs.
\subsubsection{Water sublimation in spherical envelope} 
To test this scenario for a purely spherical envelope, we constructed a model of the \irasfemtontre\ system using RADMC-3D\footnote{http://www.ita.uni-heidelberg.de/\about dullemond/software/radmc-3d/}  \citep{Dullemond:2004rf}. This code uses the method described by \citet{Bjorkman:2001gd} to calculate the dust temperature distribution in a pre-defined geometry. In this work, fifty million photon packets were propagated through the model where the infalling envelope density profile was taken from the best-fit model presented in \citet{Mottram:2013qy} and followed a power law with an exponent \mbox{$p$~=~--1.4}, where $n_{\rm{H_2}}$~=2\texpo{9}~\cmthree\ at 6.1 AU. The spherical model has a radius of 1000  AU so that it covers the region from
which the emission was analysed.\ One thousand grid points in the outflow direction were used, and the spacing between them was logarithmic, meaning that fewer grid points were used at larger radii. One thousand grid points were also used in the azimuthal direction, evenly distributed in this case. 
The dust-to-gas mass ratio was taken to be 100. A recent study towards the $\rho$~Oph~A region  showed that this ratio can vary by at least a factor of 2 \citep{Liseau:2015df}. This does not affect the results presented here, however, since the densities we used are assumptions and not actual estimates. No scattering was used here, and tests showed that this assumption does not alter our conclusions. Using anisotropic scattering will not change the distances reported here by more than 10\%.

\begin{figure*}[t!]
   \centering
   %\vspace{-1cm}
   %\hspace{-1.10cm}
    \rotatebox{0}{\includegraphics[width=1.02\textwidth]{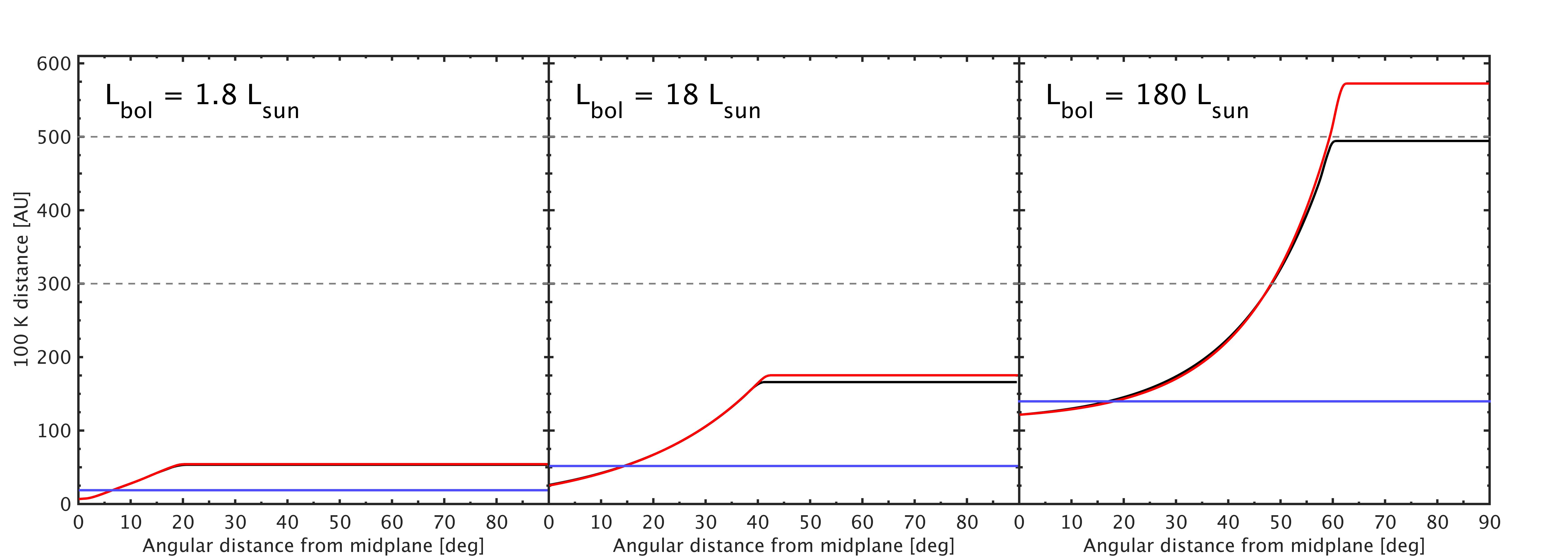}}
     % \vspace{-0.7cm}
      \caption{100 K distance from the protostar as a function of the angle to the mid-plane of the system. The curves show the distance at which the temperature drops below 100~K for three different scenarios. In the first scenario (blue), a pure envelope density profile is considered. In the second scenario (black), an outflow cone with a constant density at \expo{5}~\cmthree\
 is added to the model. The third scenario (red) is identical to the second, with the exception that the outflow density is lower, \expo{4}~\cmthree. The left panel shows the results for a protostar with \lbol~=~1.8~\lsun. The middle and right panels show the same curves when the luminosity is increased by one and two orders of magnitude, respectively. Dashed grey lines indicate the largest distance from the continuum peak for blue-shifted HDO emission (\about300 AU) and red-shifted emission (\about500~AU).}
         \label{fig:radmc}
   \end{figure*}

The results from these calculations are presented in Fig.~\ref{fig:radmc}. 
   The 100 K distance is plotted (blue lines) as a function of angular distance from the axis perpendicular to the direction of the outflow. For a spherical envelope, this is a straight line, since the temperature only depends on the radius. Three different cases were considered, one with the same luminosity as the current luminosity of the protostar, and two cases with
a luminosity increased by one and two orders of magnitude, respectively. It is clear that an envelope profile cannot explain the large extent of the HDO emission. Not even for the case when the luminosity is increased by two orders of magnitude is the 100 K radius farther away than 150 AU from the protostar.
\subsubsection{Water sublimation in spherical envelope, displaced by outflow motions}
Another possibility is that the gas has been displaced outwards by the motion of the outflow. In that case, and during the lifetime of the source, the extent of the water-emitting region would increase. Although it is difficult in this scenario to explain that the H$^{13}$CO$^+$ ring is centred on the continuum peak, we cannot exclude it from the water data alone. One major problem with this scenario, however, is that the detected highest velocities are fairly small in comparison. The highest velocity of the HDO emission is \about1~\kmpers\ with respect to the systemic velocity. Although there can be gas moving at higher velocities, the bulk part of the HDO emitting gas must move slowly. At the current velocity, this gives us a timescale of \about2300 years for the gas to reach the location of the red-shifted knot and \about1600 years to reach the farthest extent of the blue-shifted emission. These timescales are both too long compared to the estimated timescale of the most recent accretion burst \citep[100 -- 1000
years ago,][]{Jorgensen:2013lr}. Even if we (given the low signal-to-noise ratio attained for HDO) assume that this is an underestimation by a factor of two, velocities are still on the lower end. We therefore conclude that this is most likely not the main cause of the extended emission. 

\subsubsection{Water sublimation in low-density outflow}
\begin{figure}[t!]
 %  \centering
 \hspace{-0.5cm}
    \rotatebox{0}{\includegraphics[width=0.53\textwidth]{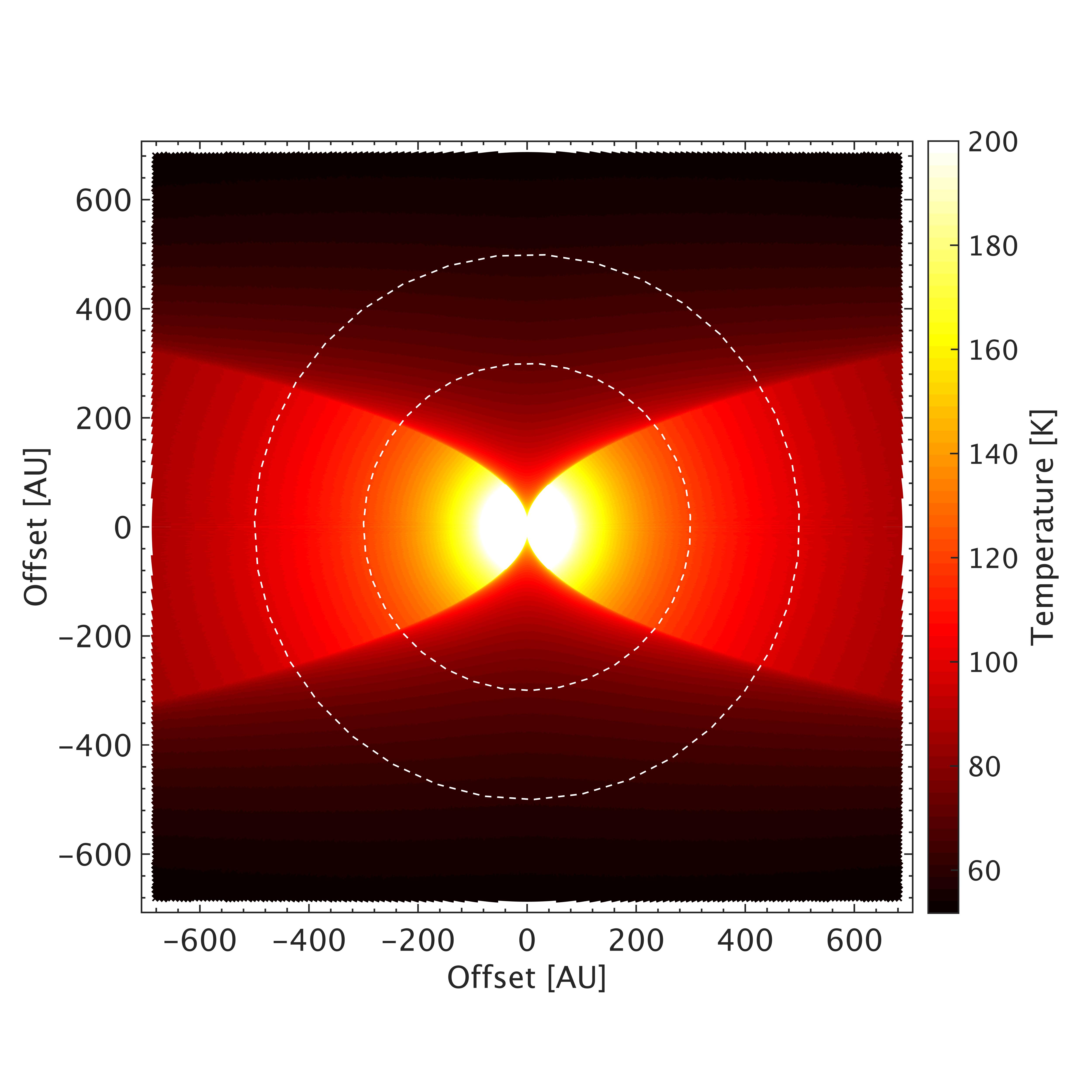}}
      \caption{Temperature map of the region when the density in the outflow cavity is constant at $n_{\rm{H_2}}$~=~\expo{5}~\cmthree\ and the luminosity of the protostar is increased by two orders of magnitude, i.e. 180~\lsun. Only the region with a temperature lower than 200 K is presented. Dashed white lines indicate the largest distance from the continuum peak for blue-shifted HDO emission (\about300 AU) and red-shifted emission (\about500 AU).}
         \label{fig:radmcmap}
   \end{figure}

Yet another possibility is that water was released from the grains at the time when the protostar underwent a luminosity burst, but that the distance out to where this occurred was changed significantly compared to what is expected for a simple power-law density profile. If densities are lower in the inner parts of the outflow cavity, then the 100~K radius can be changed significantly compared to the case where only an envelope is present. To test that, we adopted a model similar to the one presented in \citet[][Model M3]{Bjerkeli:2016lr}. The geometry of the model presented in that paper resembles the structure of the emission and the observed line profiles well, both the large-scale CO outflow and the small-scale C$_2$H emission, tracing the outflow cavity walls \citep{Jorgensen:2013lr}. The infalling envelope density profile in this case is the same as before, but we also included an outflow that is described by a wind-driven shell \citep[][their Fig. 21]{Lee:2000zr}. The details of this model can be found in \citet{Bjerkeli:2016lr}. Based on the available data towards \irasfemtontre\ we cannot estimate the density of the gas inside the outflow cone, and we therefore tested different density distributions (envelope and outflow where $n_{\rm{H_2}}$~=~\expo{5}~\cmthree, and envelope and outflow where $n_{\rm{H_2}}$~=~\expo{4}~\cmthree) to investigate how this affects the distance to which the temperature drops below 100 K.  These scenarios are both consistent with the non-detection of H$_2^{18}$O. For a temperature at 100~K and a density of \expo{5}~\cmthree,  the strength of the \htvaartonotva\ line should be a factor of \about \expo{5} lower than for the \hdoett\ line. A temperature map of the region when the outflow density is constant at $n_{\rm{H_2}}$~=~\expo{5}~\cmthree\, and the luminosity of the protostar is increased by two orders of magnitude is presented in Fig.~\ref{fig:radmcmap}. 

This figure clearly shows that if the density in the outflow cavity is lower than in the envelope at equal distances from the protostar, then the 100~K radius is indeed displaced to larger radii. A similar map was also created (but not presented here) for the case where the outflow density is even lower, $n_{\rm{H_2}}$~=~\expo{4}~\cmthree. The inclination of the outflow with respect to the plane of the sky, which has been estimated to be 20\adeg\ \citep{Oya:2014kx}, does not alter these numbers by more than 6\%.
 Owing to the lower density in the outflow, there is in these cases an angular dependence on the 100~K distance, and this is presented for the two outflow scenarios along with the envelope-only scenario in Fig.~\ref{fig:radmc}. 
 It is obvious that for the 100~K radius to be as far out as 500~AU, a high-luminosity central heating is required that would
be 10 -- 100 times of its current luminosity. 
In addition, a relatively low outflow density is required,  \nhtva~$<$~1\texpo{5}~\cmthree. 

\section{Conclusions}
From the observations presented in this paper we conclude the following:
\begin{itemize}
\item \hdoett\ at 464.925~GHz is detected towards \irasfemtontre\ and peaks at the 5$\sigma$ level (integrated emission). \htvaartonotva,
 on the other hand, is not detected, which implies a pressure lower than \about\expo{11}~\cmthree~K. The presence of water in the region where H$^{13}$CO$^+$ is not detected, is consistent with a recent burst scenario where HCO$^+$ is destroyed through reactions with water.
\item The morphology of the HDO emission shows that water is present in the outflow as well, but not at distances farther away than \about500 AU from the protostar. Although we cannot exclude the possibility that shocked regions play a role at a lower level, the observed velocities and spatial distribution are not easily reconcilable with shock chemistry. It is also unlikely that the water was released from the grains close to the protostar and transported outwards by the outflow. 
\item RADMC-3D models of a geometry containing an infalling envelope and an outflow at lower density showed that the 100~K distance from the protostar can be displaced outwards to \about500~AU. The same models show that the 100~K distance is within 150 AU from the protostar when the outflow has the same density distribution as the envelope. We therefore find it most likely that the water was released from the grains in an extended hour-glass configuration during a recent accretion burst. 
\item In addition to the targeted science lines, N$_2$H$^+$\,(5--4) and CS\,(8--7) were also detected in the mapped region within the same setup. N$_2$H$^+$ traces the outflow cavity walls and shows a rich velocity structure, while CS mainly traces the shocked regions and the central condensation.
\end{itemize}
To summarise, these data demonstrate the use of imaging molecular emission as a tracer of the evolutionary histories of embedded protostars. Although care must be taken when separating the various components of the protostellar systems (e.g. envelopes, disks, and outflows), the high angular resolution images of ALMA provide the necessary direct constraints. Studies of the molecular emission towards protostars combined with detailed modelling efforts (e.g. chemical and radiative transfer) will be an important tool in the future, in particular when larger samples of sources are systematically targeted in similar lines.
\label{section:Conclusions}

%\begin{acknowledgements}
\section*{}
      \small{\textit{Acknowledgements.}  This paper makes use of the following ALMA data:
ADS/JAO.ALMA\#2013.1.00244.S and \#2011.0.00628.S. ALMA is a partnership of ESO (representing
its member states), NSF (USA) and NINS (Japan), together with NRC
(Canada), NSC and ASIAA (Taiwan), and KASI (Republic of Korea), in
cooperation with the Republic of Chile. The Joint ALMA Observatory is
operated by ESO, AUI/NRAO and NAOJ. We also thank Ivan Marti-Vidal,
        Wouter Vlemmings and the staff at the Nordic ARC node in
        Onsala for valuable support in reducing the dataset presented in
        this paper. This research was supported by the Swedish
        Research Council (VR) through the contract 637-2013-472 to Per
        Bjerkeli. Jes J{\o}rgensen acknowledges support by a Lundbeck Foundation Junior Group Leader Fellowship as well as the European Research Council (ERC) under the European Union's Horizon 2020 research and innovation programme (grant agreement No 646908) through ERC Consolidator Grant ``S4F''. The Centre for Star and Planet Formation is funded by the Danish National Research Foundation. Daniel Harsono is funded by Deutsche Forschungsgemeinschaft Schwerpunktprogramm (DFG SPP 1385) The First 10 Million Years of the Solar System Ð a Planetary Materials Approach.}
%\end{acknowledgements}
\bibliographystyle{aa}
\bibliography{/Users/per/Dropbox/References/papers}
\end{document}